\documentstyle[twoside,fleqn,espcrc2,epsf]{article}

\newcommand{\AmS}{{\protect\the\textfont2
  A\kern-.1667em\lower.5ex\hbox{M}\kern-.125emS}}

\title{Light quark spectrum with improved gauge and fermion actions\thanks{presented 
by S.~Gottlieb}
}

\author{C.\  Bernard,\address{Department of Physics, 
Washington University, St.\ Louis, MO 63130, USA}
        T.\ DeGrand,\address{Physics Department,
University of Colorado, Boulder, CO 80309, USA}
        C.\ DeTar,\address{Physics Department,
University of Utah, Salt Lake City, UT 84112, USA}
        Steven Gottlieb,\address{Department of Physics,
Indiana University, Bloomington, IN 47405, USA}
        Urs\ M.~Heller,\address{SCRI, The Florida State University,
Tallahassee, FL 32306-4130, USA}
        J.~Hetrick,${}^{\rm a}$
	C.~McNeile,${}^{\rm c}$
        K.~Rummukainen,\address{Department of Physics,
University of Bielefeld, D--33501 Bielefeld, Germany}
        R.\ Sugar,\address{Department of Physics,
University of California, Santa Barbara, CA 93106, USA}
        D.\ Toussaint,\address{Department of Physics,
University of Arizona, Tucson, AZ 85721, USA}
        and
        M.\ Wingate,${}^{\rm b}$
} 
       
\begin{document}

\begin{abstract}
We report on a study of the light quark spectrum using an
improved gauge action and both Kogut-Susskind and Naik
quark actions.  We have studied six different lattice
spacings, corresponding to plaquette couplings ranging from
6.8 to 7.9, with five to six quark masses per coupling.
We compare the two quark actions in terms of the spectrum
and restoration of flavor symmetry.  We also compare
these results with those from the conventional action.

\end{abstract}

\maketitle

For the past few years, it has been quite popular to add additional terms to
the Wilson gauge action, or the Wilson or Kogut-Susskind quark actions in
an attempt to reduce finite lattice spacing artifacts\cite{LEPAGEREV}.  
There has been
considerable success in improving the Wilson quark action, which has errors
of order $a$, by introducing the SW clover term \cite{SW} 
to produce an action with
errors of order $a^2$.  Here we report on what may be the more challenging
problem of improving the Kogut-Susskind quark action, which only has
errors of order $a^2$ to start with.  
We have extended the study we reported on last
year to three weaker couplings \cite{MILC96}.

For the gauge fields, we use a tadpole improved three-term action that
includes the plaquette, $1\times2$ rectangle and a six link term that
corresponds to a path with steps $+x$, $+y$, $+z$, $-x$, $-y$, $-z$.
The couplings for each term are denoted $\beta_{pl}$,
$\beta_{rt}$ and $\beta_{pg}$.

The Naik quark action\cite{NAIK} extends the staggered 
quark action to include a 
three-link derivative term in addition to the usual nearest neighbor
derivative.  If the coefficients of the one and three link terms are
denoted $c_1$ and $c_3$, the Kogut-Susskind action is $c_1=1$ and $c_3=0$,
while the tree-level tadpole-improved Naik action 
is $c_1=9/(8u_0)$ and $c_3= -1/(24 u_0^3)$.  
Here $u_0$ is the tadpole improvement factor that we take to be the
fourth root of the plaquette.
The Naik action improves the free quark dispersion relation to remove order
$a^2$ errors.  Luo \cite{LUO} has identified 15 independent dimension six
lattice operators that might be added to Kogut-Susskind action to form an
order $a^2$ improved action for the interacting theory.  
Thus, going beyond tree level improvement
may require terms such as ``fat link'' interactions\cite{MILCFAT} 
and next-nearest neighbor interactions.

\begin{figure}[t]
\epsfxsize=0.99 \hsize
\epsffile{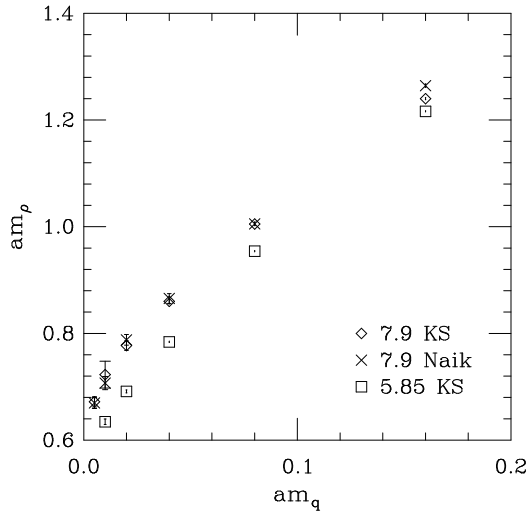}
\vspace{-28pt}
\caption{Rho mass {\it vs}.\ quark mass for standard and improved actions.}
\label{fig:rhomass}
\end{figure}

\begin{figure}[thb]
\epsfxsize=0.99 \hsize
\epsffile{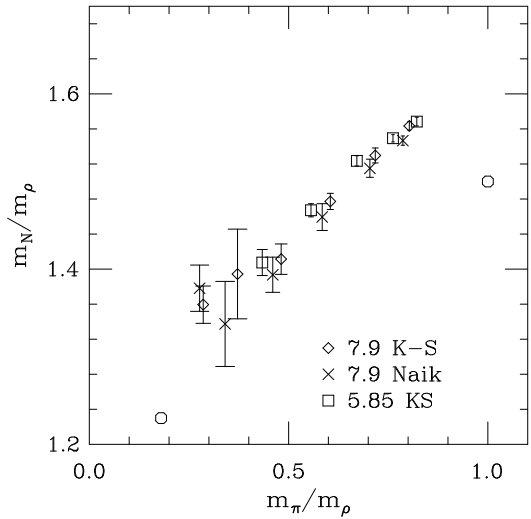}
\vspace{-28pt}
\caption{Edinburgh plot comparing the standard action with $6/g^2=5.85$ and
the improved actions with $\beta_{pl}=7.9$.}
\label{fig:edinburgh}
\end{figure}

We have three new ensembles of $16^3\times32$ lattices with $\beta_{pl}=
7.6$, 7.75 and 7.9.  For 7.6, we have 100 lattices which were provided by
the SCRI group\cite{SCRI}.  For the other couplings we have 200 lattices.  Our
previous stronger coupling work used couplings 6.8, 7.1 and 7.4, again with
200 lattice at each coupling.
We used the same volume as above, except for 7.1,
where we used $14^4\times28$.  We found that 7.4 was the
strongest coupling that gave a reasonable Edinburgh plot\cite{MILC96}, 
so we have moved
on to weaker couplings.  Results at 7.6 were not ready in time for the
conference, but are included in some of the graphs for completeness.

For each ensemble we have computed hadron propagators using five or six
quark masses differing by a factor of two.  At the strongest couplings the
hadrons are quite heavy in lattice units.  For instance at 7.1, $am_N$ is
between 2 and 3.4, while $am_\rho$ is between 1.3 and 2.2.  By 7.4, the
$\rho$ mass extrapolated to $am_q=0$ is about 1.2 and the nucleon
is 1.7.  This $\rho$ mass is comparable to $6/g^2=5.54$ for the standard
Wilson gauge action and staggered quarks.  In Fig.~1, we compare our
weakest coupling 7.9, with the standard action $6/g^2=5.85$.  We see that
5.85 has a somewhat lighter rho, and hence we infer a smaller lattice
spacing.  Thus, our improved action results correspond to a slightly
smaller range than 5.54--5.85 in terms of the standard action.  In terms
of the lattice spacing, this range is roughly 0.3--0.15 fm.

We compare Edinburgh plots for $6/g^2=5.85$ with $\beta_{pl}=7.9$ in
Fig.~2.  Despite the fact that 5.85 has a smaller lattice spacing
(as determined by the $\rho$ mass), the results for the improved gluonic
action give a nucleon to rho mass ratio that may be slightly lower.  This
improvement is actually clearer at stronger couplings as we shall see
later.

\begin{figure}[thb]
\epsfxsize=0.99 \hsize
\epsffile{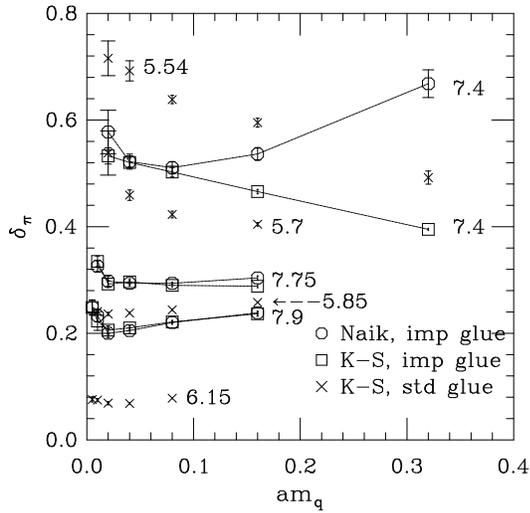}
\vspace{-28pt}
\caption{Flavor symmetry breaking variable $\delta_\pi$ {\it vs}.\ quark
mass for various actions and couplings.}
\label{fig:deltapi}
\end{figure}

We next turn to the issue of flavor symmetry restoration.  We have found
that adding ``fat link'' terms to the fermion action can result in a
substantial improvement in flavor symmetry\cite{MILCFAT}.  In Fig.~3, we display
$\delta_\pi = ( m_{\pi_2}^2 - m_\pi^2) / M_\rho^2$ {\it vs}. quark mass for
our standard staggered spectrum with $6/g^2=5.54$, 5.7, 5.85 and 6.15, and
for the improved gluonic action with either Kogut-Susskind or Naik quarks.
Except at the largest quark mass, $\beta_{pl}=7.4$ clearly has smaller
values of $\delta_\pi$ than the comparable standard coupling 5.54.
For $\beta_{pl}=7.9$, $\delta_\pi$ is smaller than the 
standard action 5.85 results,
which are the crosses between the octagons and squares for 7.75 and 7.9.
Thus, we see improvement in flavor symmetry restoration, although it seems to
be coming more from the improvement in the gluonic action than the quark
action.

\begin{figure}[thb]
\epsfxsize=0.99 \hsize
\epsffile{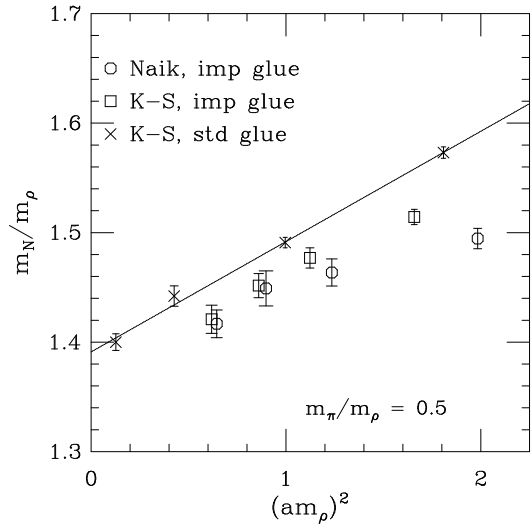}
\vspace{-28pt}
\caption{Scaling plot of nucleon to $\rho$ mass ratio for quark mass such
that $m_\pi/m_\rho=0.5$.}
\label{fig:scalingplot}
\end{figure}

For the Wilson action quenched hadron spectrum, determining the correct form
of the chiral extrapolation has been a crucial issue.  This problem is also
present for the current calculations, but there is insufficient space for a
full discussion, and our study of this issue has not yet been completed.
At the conference we showed graphs of the nucleon to rho mass ratio
extrapolated to physical quark mass {\it vs}. $am_\rho$.  Such graphs
yielded scant encouragement as to the value of the
current attempt at improvement.
In view of the difficulty in controlling the chiral extrapolation, we
display here a chiral interpolation to $m_\pi/m_\rho=0.5$.
The crosses represent the standard gluonic action.  A curve with the
expected quadratic lattice spacing dependence is drawn.  From left to right,
the couplings are 6.15, 5.85, 5.7, 5.54.  The results with improved glue
are all below the standard curve.  For the two stronger couplings,
$\beta_{pl}=7.6$ and 7.4, there is a noticeable difference between the Naik
and Kogut-Susskind actions.  However, for the two weaker couplings, 7.75 ad 7.9,
they are indistinguishable on this plot.  It is possible that by adding
terms such as suggested by Luo, one might be able to further improve the quark
action so there is a difference between the staggered and improved quark
actions even for $am_\rho \leq 1$.

We find a modest improvement to flavor symmetry restoration and the nucleon to 
rho mass ratio by using an improved gluon action with either Kogut-Susskind or 
Naik quarks when $am_\rho(m_q\approx0) \approx 1.2$.  For smaller 
lattice spacing, it is possible that adding additional terms as detailed by Luo
\cite{LUO} will lead to an action with errors of order $a^4$.  More work is
required to do precision physics with these actions 
or more refined ones.

This work was supported by the DOE and the NSF.  Computations were done at
Indiana University, PSC, SCRI and SDSC.

\end{document}